\documentclass[
twocolumn,
showpacs,
preprintnumbers,
amsmath,
amssymb,
prb,
aps,
superscriptaddress
]{revtex4-1}
\usepackage[dvipsnames,svgnames,x11names]{xcolor}

\usepackage{amsmath,amsthm,amssymb,amsfonts}

\usepackage{graphicx}
\usepackage{dcolumn}
\usepackage{bm}
\usepackage{units}%

\usepackage{tabularx}

\usepackage{tikz}
\usepackage{multirow}

\usepackage{siunitx}
\usepackage[english=american]{csquotes}
\usepackage{esint}
\usepackage{bm}
\usepackage{accents}
\usepackage{relsize}

\usetikzlibrary{calc}
\usepackage{float}
\usepackage[english,capitalise]{cleveref}

\pdfoutput=1

\begin{document}
\newlength{\LL} \LL 1\linewidth
\title{Absence of strong skew scattering in crystals with multi-sheeted Fermi surfaces}
\author{Albert~H\"onemann}
\email[E-mail: ]{albert.hoenemann@physik.uni-halle.de}
\affiliation{Institute of Physics, Martin Luther University Halle-Wittenberg,
06099 Halle, Germany}
\author{Christian~Herschbach}
\affiliation{Institute of Physics, Martin Luther University Halle-Wittenberg,
06099 Halle, Germany}
\author{Dmitry~V.~Fedorov}
\affiliation{Physics and Materials Science Research Unit, University of Luxembourg, L-1511 Luxembourg}
\author{Martin~Gradhand}
\affiliation{H.~H.~Wills Physics Laboratory, University of Bristol, Bristol BS8 1TL,
United Kingdom}
\author{Ingrid~Mertig}
\affiliation{Institute of Physics, Martin Luther University Halle-Wittenberg,
06099 Halle, Germany}
\affiliation{Max Planck Institute of Microstructure Physics, Weinberg 2,
06120 Halle, Germany}

\date{\today}

\begin{abstract}
We consider an extrinsic contribution to the anomalous and spin Hall effect in dilute alloys based on Fe, Co, Ni, and Pt hosts with different substitutional impurities. 
It is shown that a strong skew-scattering mechanism is absent in such crystals with multi-sheeted Fermi surfaces. 
Based on this finding, we conclude on the mutual exclusion of strong intrinsic and skew-scattering contributions to the considered transport phenomena. 
It also allows us to draw general conditions for materials where the anomalous Hall effect caused by the skew scattering can be achieved giant.
\end{abstract}

\maketitle
Among the phenomena caused by spin-orbit coupling (SOC), the spin Hall effect (SHE)~\cite{Dyakonov71,Hirsch99,Zhang2000,Sinova15} and the anomalous Hall effect (AHE)~\cite{Nagaosa06,Sinitsyn08,Nagaosa10} have been extensively studied during the last decade. 
The interest in the two related phenomena is caused by their promise for novel spintronics devices. 
The efficiencies of these effects can be described by the spin Hall angle (SHA) and the anomalous Hall angle (AHA), which are defined as 
\begin{align}
\alpha_{\textsf{SHE}}=\frac{\sigma^z_{yx}}{\sigma_{xx}}\qquad\text{and}\qquad
\alpha_{\textsf{AHE}}=\frac{\sigma_{yx}}{\sigma_{xx}}\ \, .
\label{eq:alphaSHEAHE}
\end{align}
Here, the spin quantization axis is chosen along the $z$ axis. 
The spin and anomalous Hall conductivity are given by $\sigma^z_{yx}$ and $\sigma_{yx}$, respectively, whereas $\sigma_{xx}$ describes the longitudinal charge conductivity~\cite{Gradhand10}.

Historically, the AHE describes a spontaneous transverse charge current in a magnetically ordered sample, whereas the SHE is usually referred to as a transverse spin current generated in nonmagnetic materials with SOC. 
Nontheless, the underlying mechanisms, i.e. intrinsic contribution~\cite{Karplus1954}, skew scattering~\cite{Smit55,Smit58} and side jump\cite{Berger1970}, are the same for both phenomena. 
Consequently, and related to \cref{eq:alphaSHEAHE}, it seems more appropriate to call the emerging transverse spin current SHE and the transverse charge current AHE. 
Still, there is no AHE but a finite SHE in nonmagnets though AHE as well as SHE can be found and utilized in magnetic materials. 
Remarkably, it is a transverse \emph{pure} spin current created by the SHE in nonmagnets, which is not the case for magnetic systems.

For practical applications, materials with large SHA and AHA are of interest.
A promising way to significantly enhance the SHE is to tune the skew scattering~\cite{Fert81,Gradhand10,Gradhand10_2,Niimi12,Herschbach12,Herschbach14,Gu15}. 
This mechanism provides the dominant contribution to the SOC-caused transverse transport in many dilute alloys~\cite{Onoda06,Lowitzer11,Niimi11,Fert11,Chadova15} and is responsible for the giant SHE found in Cu(Bi) alloys~\cite{Gradhand10_2,Niimi12}. 
However, no giant AHE caused by this mechanism was reported for bulk ferromagnets up to now. 
The absence of a strong skew-scattering contribution to the SHE is also known for Pt~\cite{Gradhand_SPIN}. 
Taking into account that platinum is close to show spontaneous magnetism, it appears possible that the magnetic order is suppressing a strong skew scattering. 
This might explain the absence of a giant AHE caused by the skew-scattering mechanism. 
However, here we show that the actual reason is related to the multi-sheeted Fermi surfaces of magnetically ordered materials.

\begin{figure}[t]
\includegraphics[width=0.8\LL]{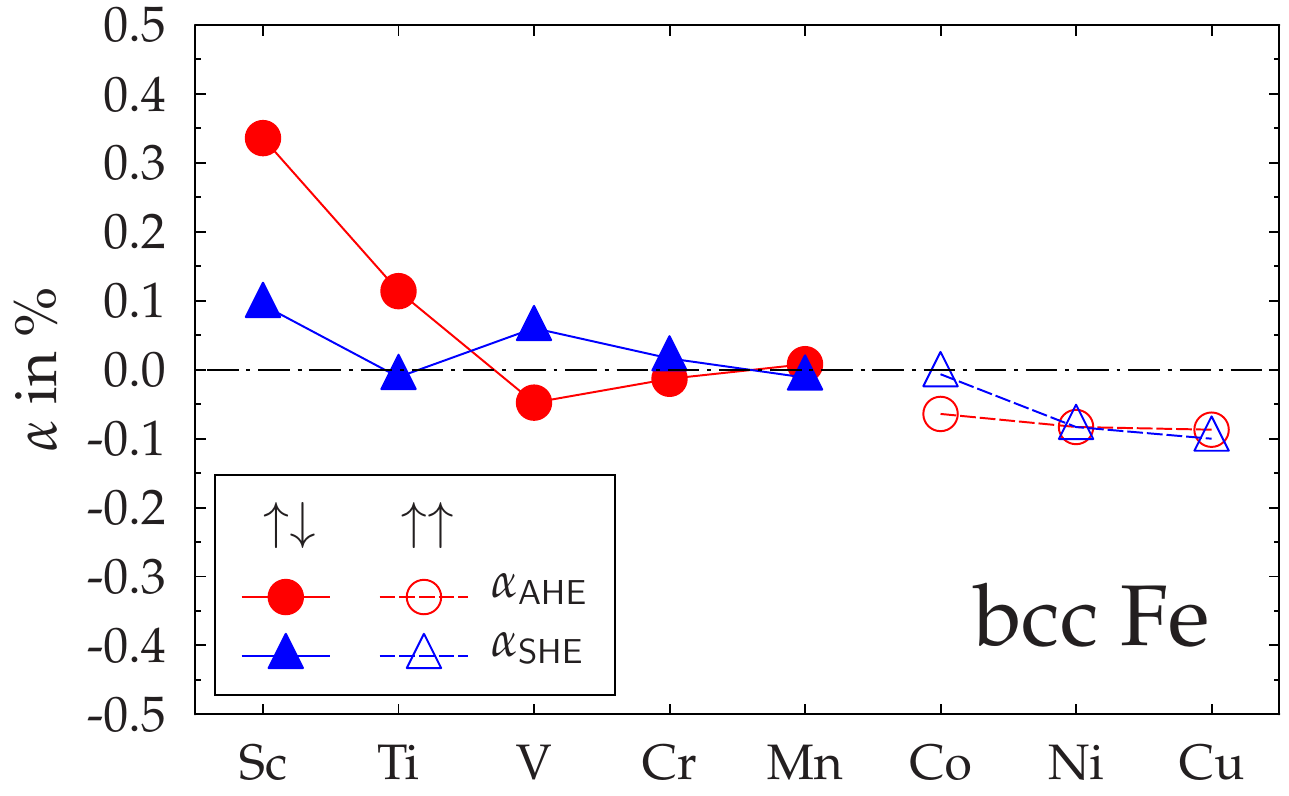}
\includegraphics[width=0.8\LL]{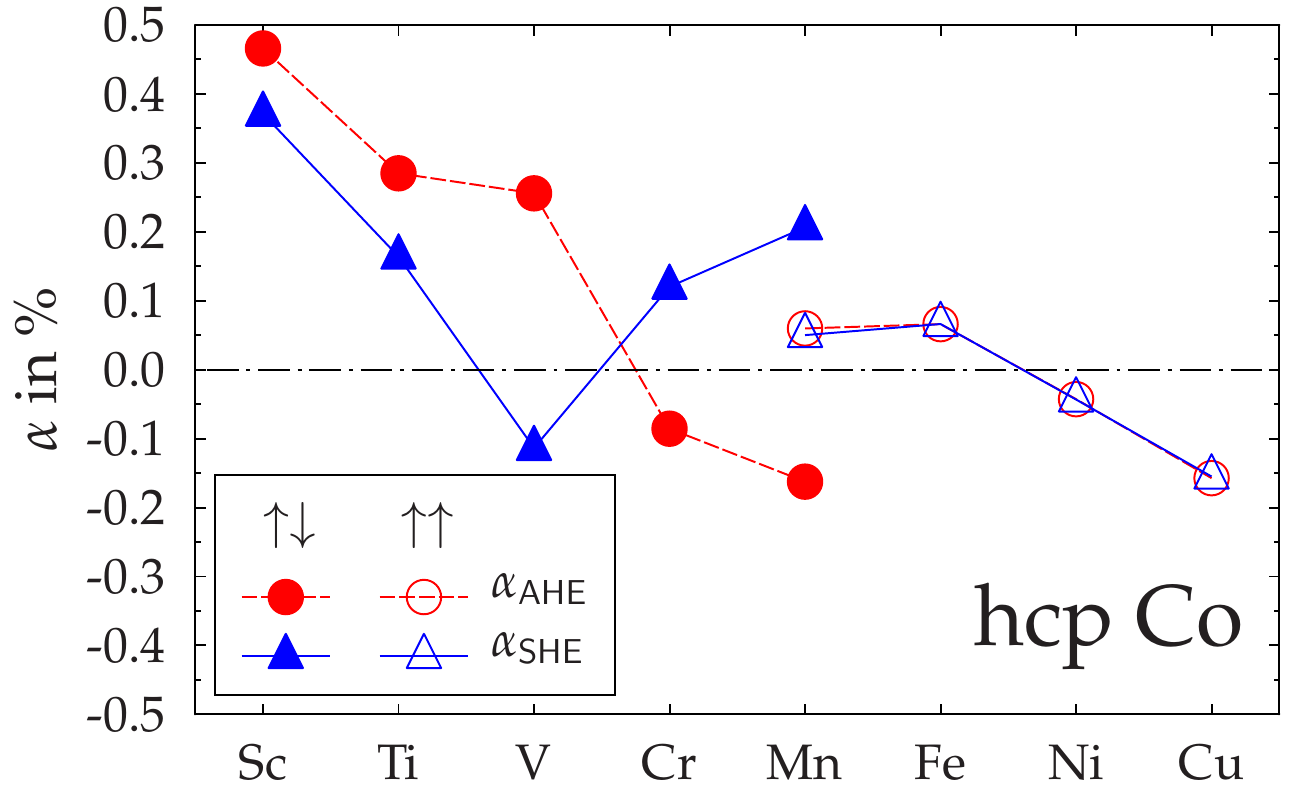}
\includegraphics[width=0.8\LL]{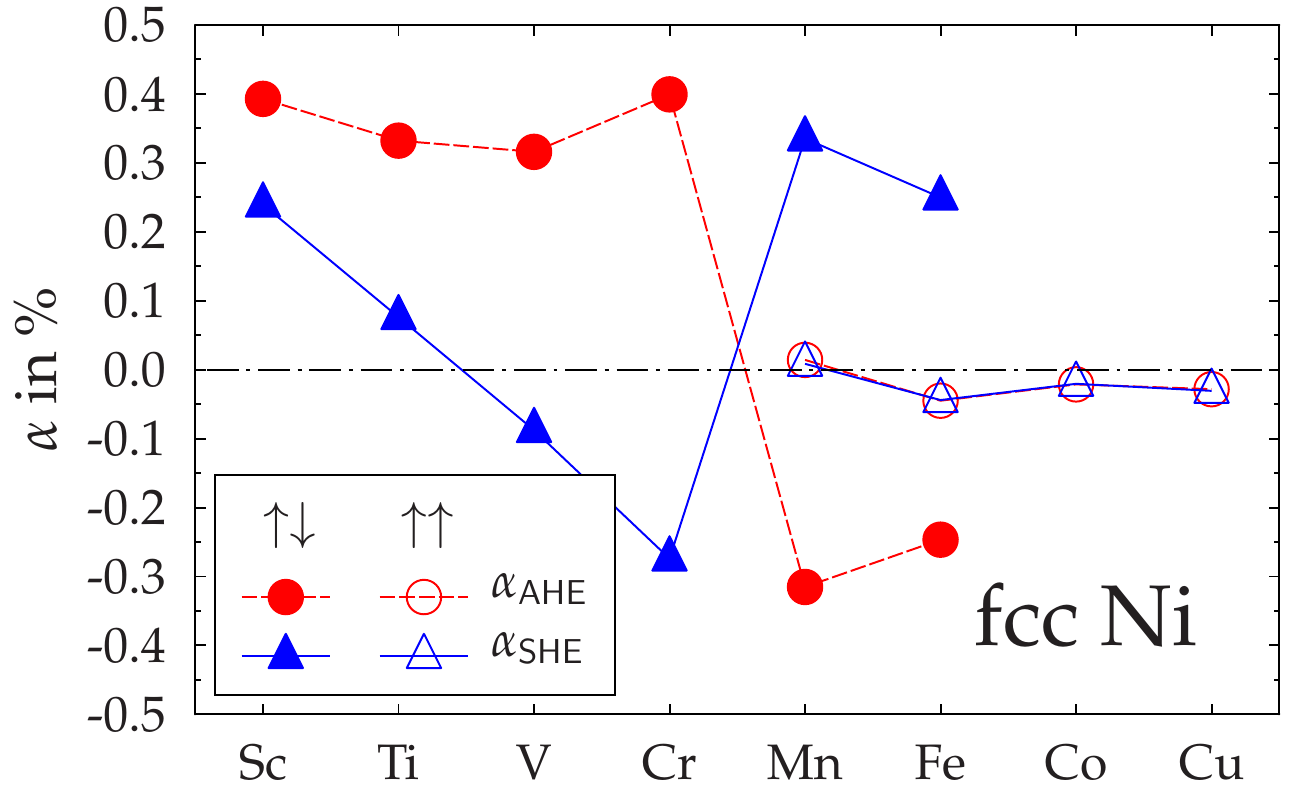}
\caption{(Color online) The anomalous Hall angle
$\alpha_\textsf{AHE}$ and the spin Hall angle $\alpha_\textsf{SHE}$
caused by $3d$-impurities in the bcc Fe, hcp Co, and fcc Ni hosts.
The filled and open symbols correspond to the antiparallel ($\uparrow\downarrow$)
and parallel ($\uparrow\uparrow$) alignment of the impurity magnetic moment with
respect to the host magnetization, respectively.}
\label{img:3d_impurities}
\end{figure}

\begin{figure}[t]
\includegraphics[width=0.8\LL]{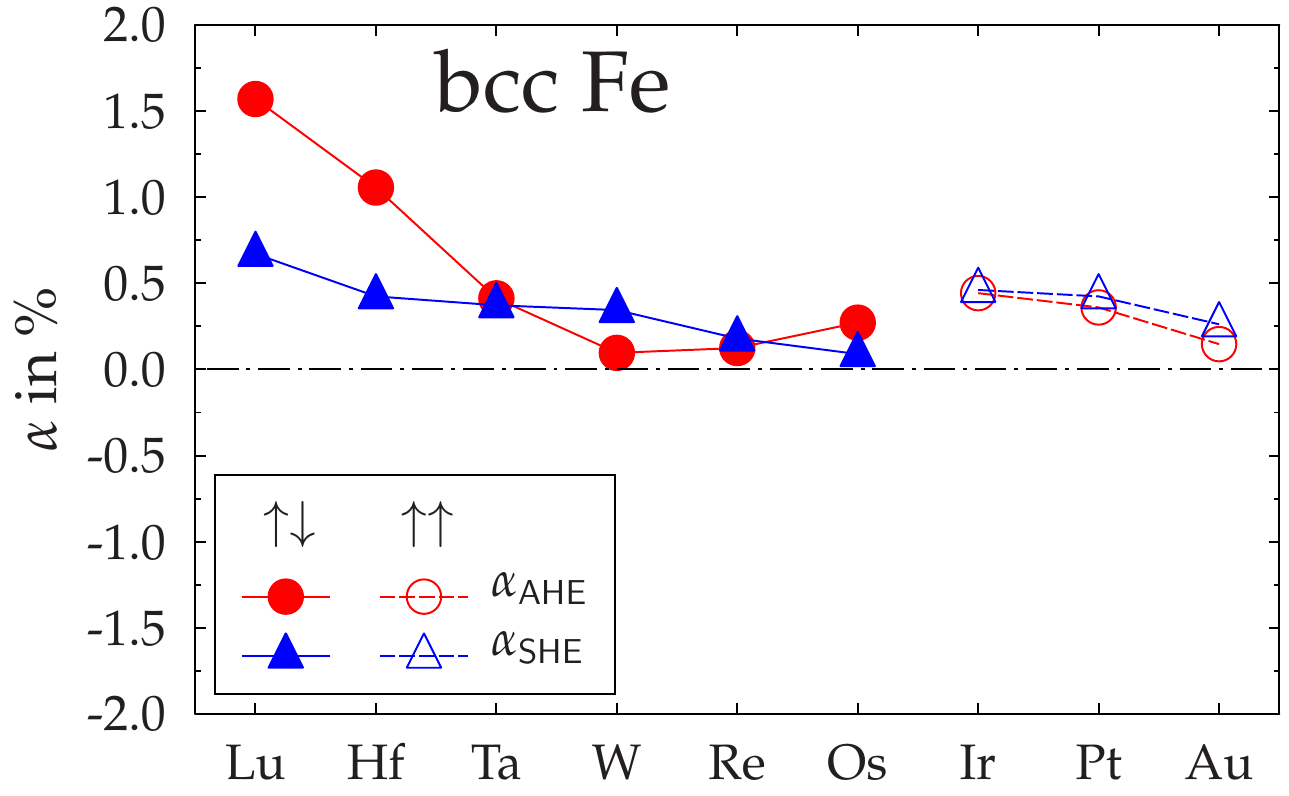}
\includegraphics[width=0.8\LL]{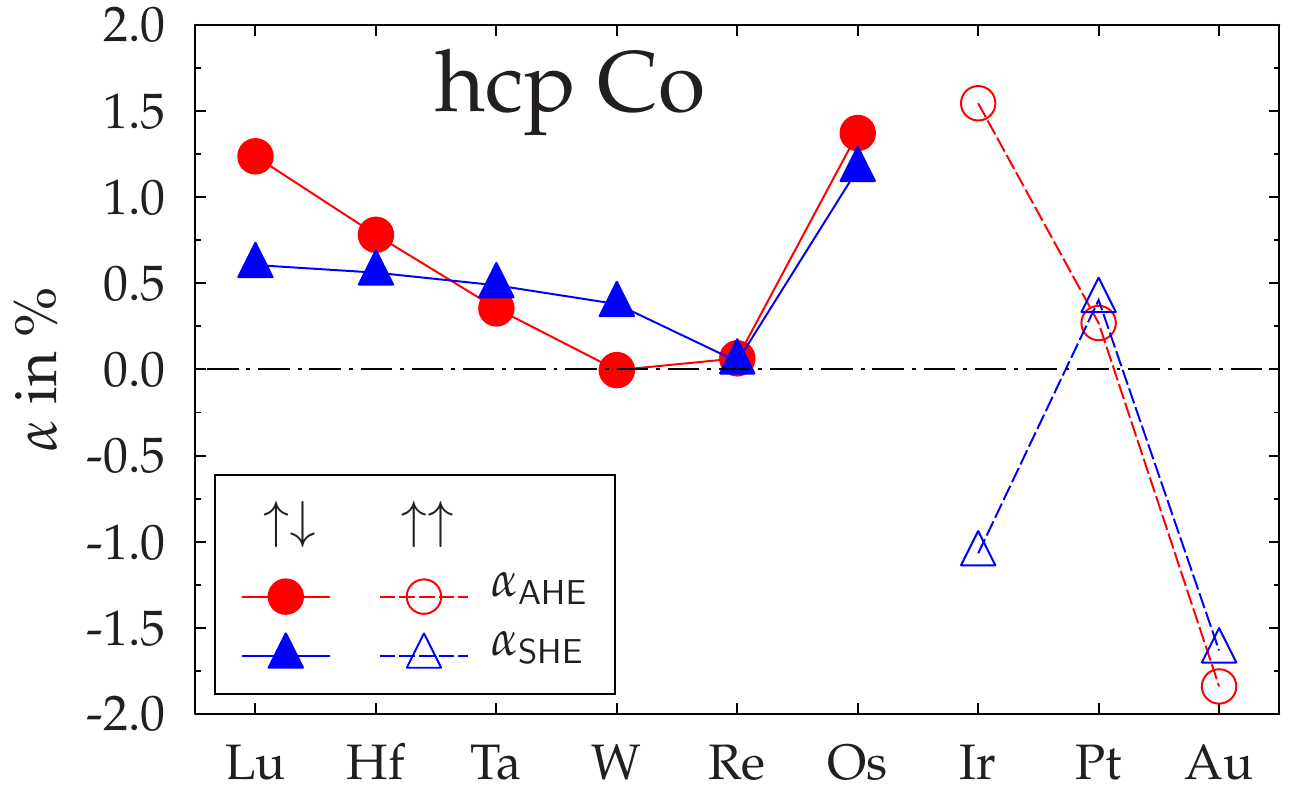}
\includegraphics[width=0.8\LL]{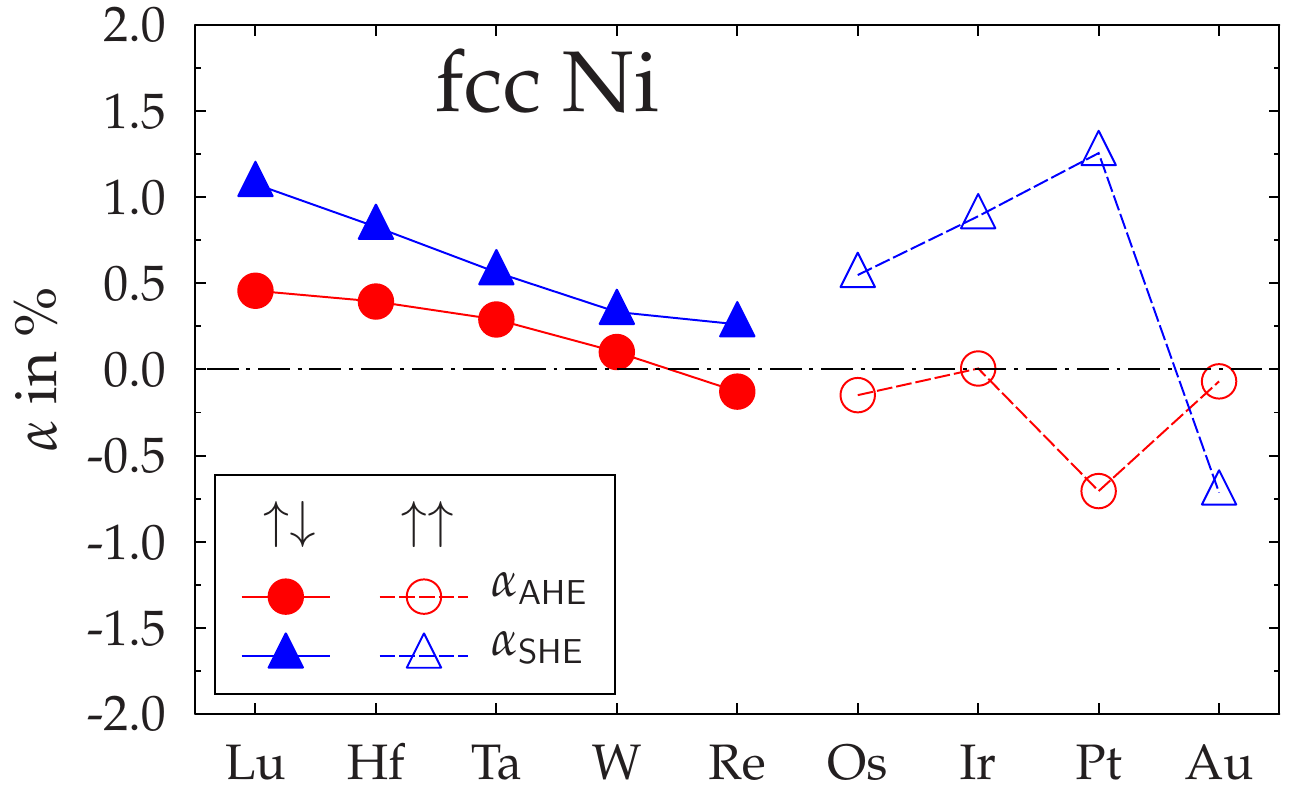}
\caption{(Color online) The anomalous Hall angle
$\alpha_\textsf{AHE}$ and the spin Hall angle $\alpha_\textsf{SHE}$
caused by $5d$-impurities in the bcc Fe, hcp Co, and fcc Ni hosts.
The filled and open symbols correspond to the antiparallel ($\uparrow\downarrow$)
and parallel ($\uparrow\uparrow$) alignment of the impurity magnetic moment with
respect to the host magnetization, respectively.}
\label{img:5d_impurities}
\end{figure}

We perform a detailed first-principles study of the AHE and SHE in dilute alloys based on
Fe, Co, Ni, and Pt hosts with different impurities. The parameters of our calculations are given in the
Supplemental Material~\cite{Supplementary}. We analyze the obtained results in comparison to Cu as host material.
The electronic structure of the hosts as well as of the impurity systems are calculated by the 
relativistic Korringa-Kohn-Rostoker Green's function method within the framework of density functional
theory\cite{GradhandKKR2009}. Their transport properties are described within
the semiclassical Boltzmann approach~\cite{Mertig99,Gradhand10,Zimmermann14} providing the longitudinal
charge conductivity as well as the skew-scattering contribution to both the spin and anomalous Hall conductivities.

\begin{figure}[t]
\includegraphics[width=0.9\LL]{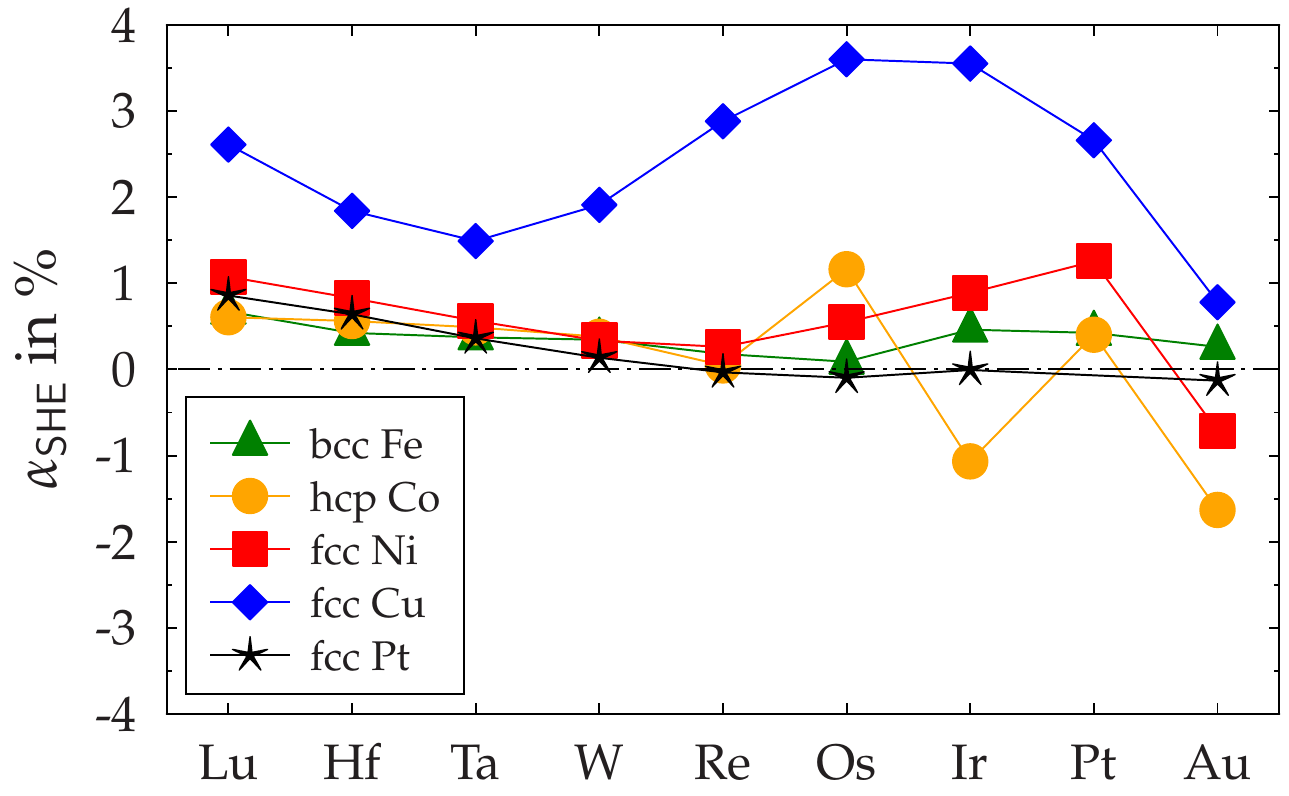}
\caption{(Color online) The spin Hall angle
$\alpha_\textsf{SHE}$ caused by $5d$-impurities in the bcc Fe, hcp Co, fcc Ni, fcc Cu, and fcc Pt hosts.}
\label{img:5d_in_all}
\end{figure}

We start our discussion by considering $3d$ and $5d$ impurities in the investigated magnetic hosts.
The corresponding results are shown in \cref{img:3d_impurities,img:5d_impurities}, respectively.
The impurity magnetic moment obtained as a self-consistent solution of the corresponding impurity problem has
either antiparallel ($\uparrow\downarrow$) or parallel ($\uparrow\uparrow$) alignment relative to
the host magnetization for impurity atoms at the beginning and the end of the $3d$ series, respectively.
More detailed information is provided by the Supplemental Material~\cite{Supplementary}.
For Mn in Co as well as Mn and Fe in Ni, both magnetic solutions can be stabilized, with only one of them being energetically preferable~\cite{Supplementary}. Nevertheless, it is interesting to consider
the transport properties for both cases. 
Such double solutions are absent in the case of $5d$ impurities, since they only show induced magnetic moments.

In order to analyze the charge and spin conductivities stemming from the Boltzmann approach, we apply the two-current model~\cite{Mott36}
\begin{align}
\hat{\sigma} = \hat{\sigma}^+ + \hat{\sigma}^-\qquad\text{and}\qquad
\hat{\sigma}^z = \hat{\sigma}^+ - \hat{\sigma}^-\ ,
\label{eq:Mott}
\end{align}
where $\hat{\sigma}^+$ and $\hat{\sigma}^-$ are the spin-resolved conductivities. Throughout the paper,
we will use the names \lq\lq spin-up\rq\rq\ (\lq\lq $+$\rq\rq ) and \lq\lq spin-down\rq\rq\ (\lq\lq $-$\rq\rq ) for
the majority and minority electrons, respectively. For the Hall components of the conductivities, we can write~\cite{Zimmermann14}
\begin{align}
\sigma_{yx}^\pm = \frac{1}{2}(\sigma_{yx} \pm \sigma_{yx}^z) = \frac{1}{2}\sigma_{xx}(\alpha_{\textsf{AHE}} \pm \alpha_{\textsf{SHE}})\ .
\label{eq:two_current}
\end{align}
Based on \cref{eq:two_current}, we gain insight into the microscopic mechanisms governing the results shown
in \cref{img:3d_impurities,img:5d_impurities}. First of all, we consider the situation when
$\alpha_\textsf{SHE} \approx \alpha_\textsf{AHE}$ as caused by Mn, Fe, Ni, and Cu impurities
in hcp Co. Following \cref{eq:two_current}, in this case the transverse transport is solely provided by
the \lq\lq spin-up\rq\rq\ channel.
The opposite occurs when $\alpha_\textsf{SHE} \approx -\alpha_\textsf{AHE}$,
as is present in case of the antiparallel configuration for Mn impurities in the three magnetic hosts.
This implies that the contribution of the \lq\lq spin-up\rq\rq\ channel is negligible and the transverse
transport is merely provided by the \lq\lq spin-down\rq\rq\ electron states. 

Commonly, only the AHE is analyzed in magnetic materials due to its easier measurement in comparison
to the SHE. Nevertheless, spin currents created by the SHE in ferromagnets are of a high interest for spintronics
technology. Figures~\ref{img:3d_impurities} and \ref{img:5d_impurities} show that, for most systems, the AHE is stronger than
the SHE. However, there are systems, like 5$d$ impurities in Ni, where the SHA is significantly larger than the AHA.
These results show that, generally, there is no correlation between the two phenomena and one needs to study them
independently. In addition, it is also worth mentioning that parallel and antiparallel alignment of the impurity
magnetic moment with respect to the host magnetization provide strongly different results. This can be
used for a simple experimental proof of an energetically preferable magnetic configuration by means of the anomalous
Hall measurements. 

\begin{figure}[t!]
\includegraphics[width=0.9\LL]{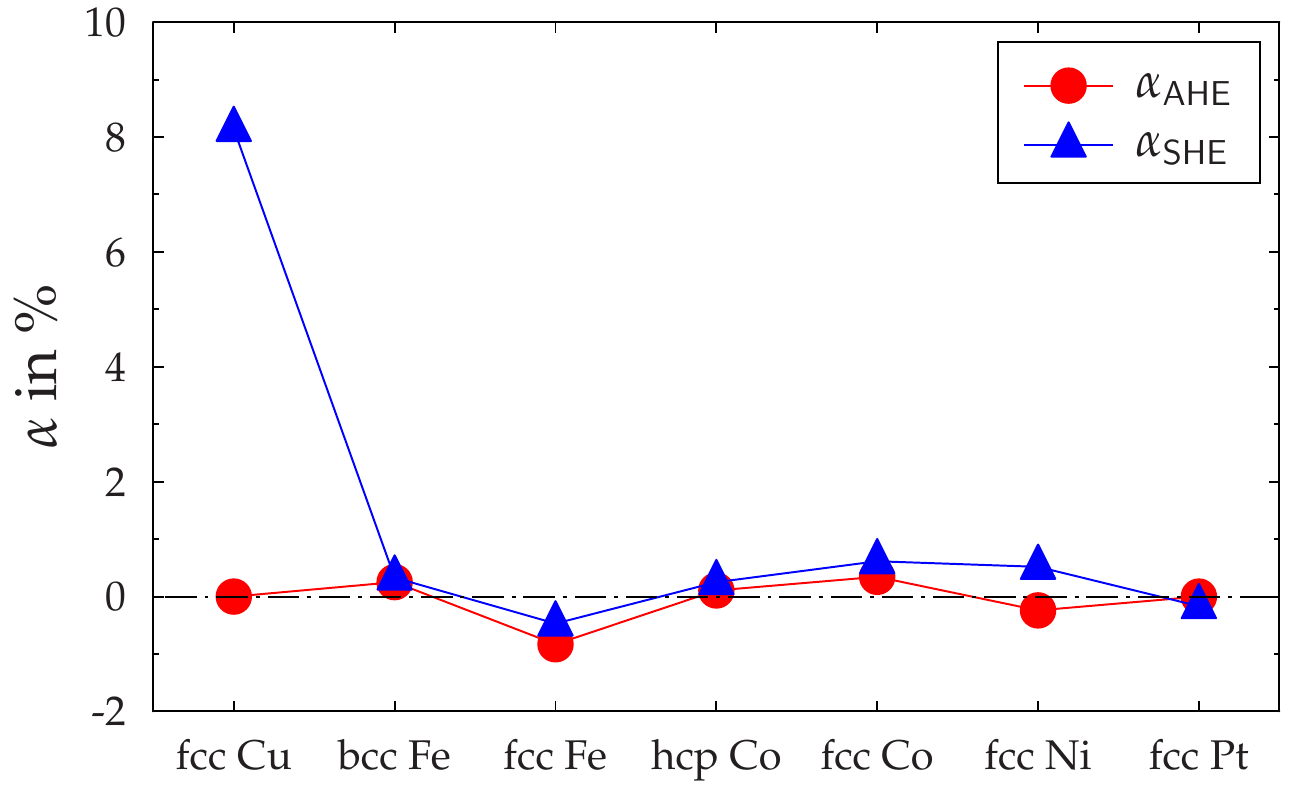}
\caption{(Color online) The anomalous Hall angle
$\alpha_\textsf{AHE}$ and the spin Hall angle $\alpha_\textsf{SHE}$
caused by Bi impurities in the Cu, Fe, Co, Ni, and Pt hosts.}
\label{img:Bi_in_hosts}
\end{figure}

Figure~\ref{img:3d_impurities} shows that in the case of $3d$ impurities the magnitude of neither AHA nor SHA exceeds
\SI{0.5}{\percent}, which indicates a weakness of the investigated transport phenomena in the considered systems. The heavier
$5d$ impurities increase the effects due to their stronger SOC. However, even in this case, the magnitudes of $\alpha_\textsf{AHE}$
and $\alpha_\textsf{SHE}$ do not exceed \SI{2}{\percent} and they are mostly below \SI{1}{\percent}. In comparison to much larger
values for corresponding SHAs in the Cu host~\cite{Johansson14}, the skew-scattering mechanism seems to be suppressed by magnetic order. 
To investigate this point we show $\alpha_\textsf{SHE}$ obtained for $5d$
impurities in magnetic Fe, Co, and Ni hosts in comparison to nonmagnetic Cu and Pt hosts (see \cref{img:5d_in_all}).
In the case of Cu
the effect is significantly increased with respect to all other hosts including Pt. This becomes even more pronounced
for the SHA and AHA caused by Bi impurities in the considered hosts, as shown in \cref{img:Bi_in_hosts}. For the Cu host,
the so-called \emph{giant SHE} is present~\cite{Gradhand10_2,Niimi12,Fedorov13} related to the SHA of about \SI{8}{\percent}.
By contrast, all other hosts, including the nonmagnetic Pt, show results with drastically reduced values of both the SHA and AHA. Therefore,
magnetic order in itself cannot be the origin of the weak skew scattering.

In order to understand the difference between Fe, Co, Ni, and Pt on one hand and Cu on the other it is instructive to analyze their multi-sheeted Fermi surfaces.
Due to the unfilled $d$ shell in all but Cu, the host crystals have several bands at the Fermi level.
For Fe, Co, and Ni, this drives their magnetic order. By contrast, Cu has just one Fermi surface sheet.

To show whether this is exactly the reason for the observed difference in the strength of the skew-scattering mechanism, we perform
auxiliary calculations. For them, each band is considered separately by excluding interband transitions, similar to the approach of
Ref.~\onlinecite{Herschbach12}. To this end, we write the total microscopic transition probability~\cite{Gradhand10,Mertig99} as
\begin{align}
P_{\bm{k}'\nu'\leftarrow\bm{k}\nu}&=
P_{\bm{k}'\nu'\leftarrow\bm{k}\nu}\delta_{\nu\nu'}
+P_{\bm{k}'\nu'\leftarrow\bm{k}\nu}(1-\delta_{\nu\nu'}) \, ,
\label{eq:PkkDecomposition}
\end{align}
where the two terms on the right hand side describe intraband ($\nu=\nu'$) and interband transitions ($\nu\neq\nu'$), respectively.
Allowing only for intraband transitions, we solve the linearized Boltzmann equation~\cite{Gradhand10,Mertig99} for each band
separately and calculate the band-restricted spin and charge conductivity tensors denoted by $\widetilde{\sigma}^{z\nu}$ and
$\widetilde{\sigma}^\nu$. The corresponding spin and anomalous Hall angles are given by
\begin{align}
\widetilde{\alpha}_\mathsf{SHE}^\nu=\frac{\widetilde{\sigma}_{yx}^{z\nu}}{\widetilde{\sigma}_{xx}^\nu}
\quad\text{and}\quad
\widetilde{\alpha}_\mathsf{AHE}^\nu=\frac{\widetilde{\sigma}_{yx}^\nu}{\widetilde{\sigma}_{xx}^\nu}\,,
\label{eq:AHASHAWOIB}
\end{align}
respectively. 
If both intraband and interband transitions are considered, the corresponding band-restricted spin and anomalous Hall angles are given by 
\begin{align}
\alpha^\nu_\mathsf{SHE}=\frac{\sigma_{yx}^{z\nu}}{\sigma_{xx}^{\nu}}
\quad\text{and}\quad
\alpha^\nu_\mathsf{AHE}=\frac{\sigma_{yx}^\nu}{\sigma_{xx}^{\nu}}
\label{AHASHAWIB}
\end{align}
with $\sigma^{z\nu}$ and $\sigma^{\nu}$ as the band-resolved spin and charge conductivities.

\begin{figure*}[t]
\includegraphics[width=2.0\LL]{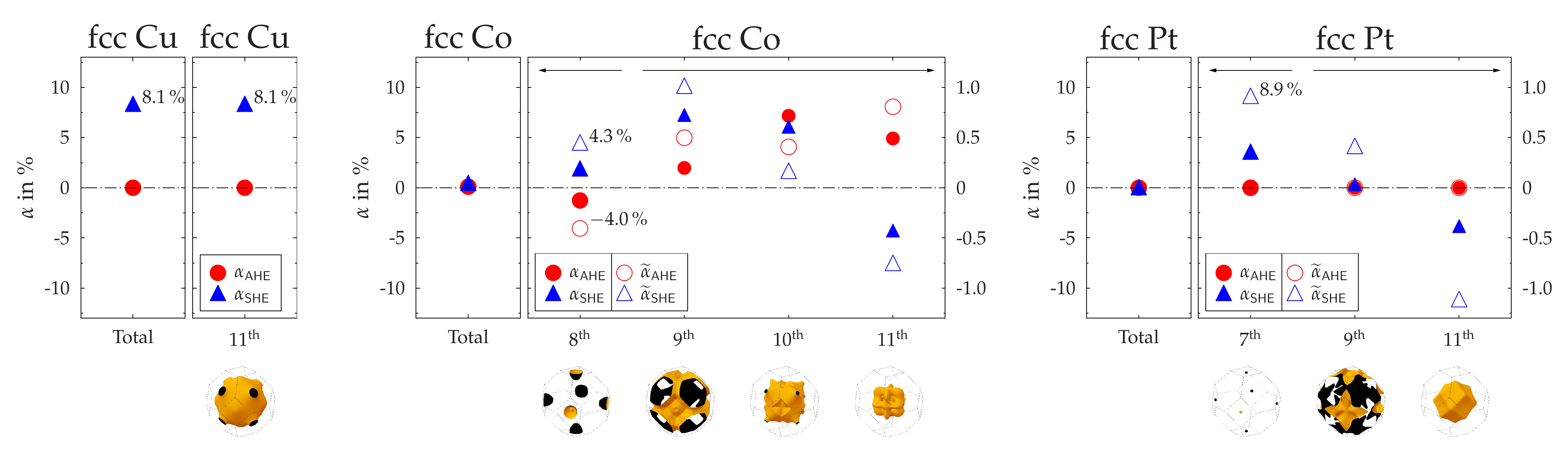}
\caption{(Color online) The band-restricted spin and anomalous Hall angles caused by Bi impurities in fcc Cu, Co, and Pt hosts.
Open symbols represent results that neglect interband transitions according to \cref{eq:AHASHAWOIB}, whereas full symbols show results that include
intraband and interband transitions, following \cref{AHASHAWIB}.}
\label{img:Band_resolved}
\end{figure*}

The results of such artificial calculations for fcc Co and Pt are shown in \cref{img:Band_resolved}. 
Basically, excluding interband scattering leads to significantly increased effects.
Additionally, for both hosts at least one band provides a giant effect. 
In the case of Co host it is the 8th band, for which Bi impurities cause the values $\SI{4.3}{\percent}$ and $\SI{-4.0}{\percent}$ for the SHA and AHA, respectively.
For the Pt host, the 7th band delivers $\alpha_\textsf{SHE} = \SI{8.9}{\percent}$ which is even slightly larger than the giant SHA obtained for the Cu host. 
However, the influence of other bands eliminates the giant effect if the total $\alpha_\textsf{SHE}$ is considered for Pt. 
This situation is similar to our study of two-dimensional systems~\cite{Herschbach12}, where it was found that interband scattering significantly reduces the skew-scattering contribution to the SHE.
Following our findings we conclude that strong skew-scattering is suppressed in crystals with multi-sheeted Fermi surfaces. However, as well known, the multi-band electronic structure is required for a large intrinsic contribution to the AHE and SHE~\cite{Fang03,Guo08}. This suggests that strong intrinsic and giant skew-scattering contributions are mutual exclusive.

Based on the obtained results, we can predict systems with a giant AHE due to the skew-scattering mechanism. 
The best candidates would be materials with properties of semimetals, where one spin channel builds a band gap
whereas another one has predominantly states of $s$ character at the Fermi level. This should provide the host Fermi surface
similar to that of copper, but with one spin channel only. According to our findings, a corresponding material can possess
a strong AHE. Further theoretical as well as experimental investigations are desirable to prove our prediction.

In summary, we have investigated the spin and anomalous Hall effect caused by the skew-scattering mechanism in dilute magnetic
alloys based on Fe, Co, and Ni. A simultaneous consideration of both the charge and spin conductivity allows for the identification
of the spin channel dominating the longitudinal as well as transverse transport. The performed study provides a detailed insight into
the skew scattering, which shows that the related mechanism is generally suppressed in crystals with multi-sheeted Fermi surfaces.
This explains the situation with a relatively week skew scattering observed in magnetic materials. Nonetheless, the SHE may become
significant even in ferromagnets with a weak AHE. We also propose a route to search for magnetic materials with a giant AHE caused
by skew scattering.

\begin{acknowledgments}
This work was partially supported by the Deutsche Forschungsgemeinschaft (DFG) via SFB 762.
In addition, M.G. acknowledges financial support from the Leverhulme Trust via an Early Career
Research Fellowship (ECF-2013-538).
\end{acknowledgments}

%

\end{document}